\documentclass[12pt]{article}
\usepackage{epsfig}
\usepackage{amssymb}
\usepackage{amsmath}
\usepackage{amsfonts}
\usepackage{graphicx}
\usepackage{mathrsfs}
\usepackage[dvips]{color}
\usepackage{multirow}

\newcommand{\R}{\mathbb{R}}
\newcommand{\C}{\mathbb{C}}

\newcommand{\bM}{\mathbf{M}}

\newcommand{\bS}{\mathbf{S}}

\newcommand{\bsigma}{\boldsymbol{\sigma}}

\newcommand{\cN}{\mathcal{N}}

\newcommand{\cP}{\mathcal{P}}

\newcommand{\cR}{\mathcal{R}}

\newcommand{\cT}{\mathcal{T}}

\newcommand{\be}{\begin{equation}}
\newcommand{\ee}{\end{equation}}
\newcommand{\bea}{\begin{eqnarray}}
\newcommand{\eea}{\end{eqnarray}}

\newcommand{\ed}{\end{document}}

\newcommand{\bi}{\begin{itemize}}
\newcommand{\ei}{\end{itemize}}
\newcommand{\xto}{\Rightarrow}

\newcommand{\bce}{\begin{center}}
\newcommand{\ece}{\end{center}}

\newcommand{\sW}{\mathscr{W}}

\begin{document}

\title{Generalized Unitarity Relation for Linear Scattering Systems in One Dimension}

\author{Ali~Mostafazadeh\thanks{E-mail address: amostafazadeh@ku.edu.tr}~
\\ Departments of Mathematics and Physics, 
Ko\c{c} University,\\ 34450 Sar{\i}yer,
Istanbul, Turkey}

\date{ }
\maketitle

\begin{abstract}
We derive a generalized unitarity relation for an arbitrary linear scattering system that may violate unitarity, time-reversal invariance, $\cP\cT$-symmetry, and transmission reciprocity. 




\end{abstract}

\section{Introduction}

The scattering phenomenon defined by a real scattering potential $v(x)$ through the time-independent Schr\"odinger equation,
	\be
	-\psi''(x)+v(x)\psi(x)=k^2\psi(x),
	\label{sch-eq}
	\ee
satisfies the unitarity relation:
	\be
	|R^{l/r}(k)|^2+|T^{l/r}(k)|^2=1,
	\label{unitarity-rel}
	\ee
where $R^{l/r}(k)$ and $T^{l/r}(k)$ are respectively left/right reflection and transmission amplitudes. The latter determine the asymptotic behavior of the scattering solutions of (\ref{sch-eq}) according to
	\begin{align}
	& \psi_l(k,x)\to\left\{\begin{array}{ccc}
    	 \cN_+(k)\left[e^{ikx}+{R}^{l}(k)\,e^{-ikx}\right] & {\rm for}& x\to-\infty,\\
    	 \cN_+(k){T}^l(k)\,e^{ikx} & {\rm for}& x\to+\infty,
    	\end{array}\right.
	\label{psi-L}\\[6pt]
	& \psi_r(k,x)\to\left\{\begin{array}{ccc}
    	 \cN_-(k){T}^r(k)\,e^{-ikx} & {\rm for}& x\to-\infty,\\
    	 \cN_-(k)\left[e^{-ikx} + {R}^r(k)\,e^{ikx}\right]  & {\rm for}& x\to+\infty.
    	\end{array}\right.
    	\label{psi-R}
    	\end{align}
These respectively correspond to scattering setups where left-/right-incident waves of amplitude $\cN_{+/-}(k)$ are scattered by the potential $v(x)$.  

For a real scattering potential, one can show that \cite{jpa-2014c}
	\bea
	|R^l(k)|&=&|R^r(k)|,
	\label{R=R}\\
	T^l(k)&=&T^r(k).
	\label{recip}
	\eea
Therefore the unitarity relation takes the form
	\be
	|R^{l/r}(k)|^2+|T(k)|^2=1,
	\label{unitarity}
	\ee
where $T(k)$ stands for the common value of $T^l(k)$ and $T^r(k)$. 

Reciprocity in transmission (\ref{recip}) turns out to be a universal feature of all real and complex scattering potentials \cite{ahmed,jpa-2009}. To see this we first recall that the Wronskian of any pair of solutions $\psi_{1,2}(x)$ of (\ref{sch-eq}), i.e., 
	\[W[\psi_1(x),\psi_2(x)]:=\psi_1(x)\psi_2'(x)-\psi_1'(x)\psi_2(x),\] 
is independent of $x$. If we compute $W[\psi_l(x),\psi_r(x)]$ for $x\to -\infty$ and $x\to+\infty$, we respectively find $2ik/T^l(k)$ and $2ik/T^r(k)$. The fact that these must be equal to the same constant implies (\ref{recip}) for $k\neq 0$. This is actually the one-dimensional realization of the celebrated reciprocity theorem which is for example proven for real potentials in Ref.~\cite{landau-lifshitz}.

Unlike (\ref{recip}), (\ref{R=R}) is violated by generic complex scattering potentials. A striking demonstration of this fact is the existence of unidirectionally reflectionless complex potentials \cite{lin}. These are potentials whose reflection amplitudes fulfil either $R^l(k)=0\neq R^r(k)$ or $R^r(k)=0\neq R^l(k)$ for some $k\in\R^+$. It turns out that these conditions are invariant under the combined action of parity and time-reversal transformation ($\cP\cT$), where $\cT\psi(x):=\psi(x)^*$ and $\cP\psi(x):=\psi(-x)$ respectively define the parity and time-reversal transformations \cite{pra-2013a}. This in turn makes $\cP\cT$-symmetric potentials\footnote{These are potentials that satisfy $v(-x)^*=v(x)$.} the principal examples of unidirectionally reflectionless potentials. This together with the interesting properties of their spectral singularities \cite{ss-review} have made $\cP\cT$-symmetric scattering potentials a focus of intensive research activity during the past decade \cite{konotop-review}. 

Among the outcomes of the research done in the subject is the discovery of the following generalization of the unitarity relation (\ref{unitarity}) for $\cP\cT$-symmetric potentials \cite{ge}:
	\be
	|T(k)|^2\pm |R^l(k)R^r(k)|=1.
	\label{pseudo-unitary}
	\ee
Another curious observation is that reflection and transmission amplitudes of $\cP\cT$-symmetric scattering potentials satisfy 
	\begin{align}
	&|R^l(-k)|=|R^r (k)|, && |T(-k)|=|T(k)|.
	\label{PT-RR-TT}
	\end{align}
These were initially conjectured in \cite{ahmed-2012} based on evidence provided by the study of a complexified Scarf II potential. They were subsequently proven as immediate consequences of  the following identities that hold for $\cP\cT$-symmetric scattering potentials \cite{jpa-2014c}.
	\begin{align}
	&R^{l/r}(-k)=-e^{2i\tau(k)}R^{r/l} (k), && T(-k)=T(k)^*,
	\label{PT-RR-TT}
	\end{align}
where $e^{i\tau(k)}:=T(k)/|T(k)|$. In view of the second of these equations, we can write the first in the form
	\be
	R^{l/r}(-k)T(-k)+R^{r/l}(k)T(k)=0.
	\ee

The analysis leading to the proof of (\ref{PT-RR-TT}) also reveals that the reflection and transmission amplitudes of both real and $\cP\cT$-symmetric scattering potentials fulfil \cite{jpa-2014c}
	\be
	R^{l/r}(k)R^{l/r}(-k)+|T(k)|^2=1.
	\label{id-1}
	\ee
It is not difficult to see that this reduces to (\ref{unitarity}) and (\ref{pseudo-unitary}) for real and $\cP\cT$-symmetric potentials, respectively.

The purpose of the present article is to establish a generalization of (\ref{id-1}) that holds for every linear scattering system, even those that are not defined by a local potential \cite{muga,muga2}.

\section{General scattering systems in one dimension}

Consider a wave equation in $1+1$ dimensions that admits time-harmonic solutions: $e^{-i\omega t}\psi(x)$, where $\psi:\R\to\C$ solves a time-independent wave equation, 
	\be
	\sW[\psi,x]=0.
	\label{we}
	\ee
This equation, which may be nonlocal or even nonlinear, defines a meaningful scattering phenomenon if for $x\to\pm\infty$ its solutions tend to those of
	\be
	-\psi''(x)=k^2\psi(x).
	\label{free-wave}
	\ee 
In other words, solutions of (\ref{we}) satisfy the asymptotic boundary conditions:
	\bea
	\psi(x)&\to& A_-(k)e^{ikx}+B_-(k)e^{-ikx}~~{\rm for}~~x\to-\infty,
	\label{pw-m}\\
	\psi(x)&\to& A_+(k)e^{ikx}+B_+(k)e^{-ikx}~~{\rm for}~~x\to+\infty,
	\label{pw-p}
	\eea
where $A_\pm$ and $B_\pm$ are complex-valued coefficient functions. We call the $2\times 2$ matrices $\bM(k)$ and $\bS(k)$ satisfying 
	\begin{align}
	&\bM(k)\left[\begin{array}{c} A_-(k)\\ B_-(k)\end{array}\right]=
	\left[\begin{array}{c} A_+(k)\\ B_+(k)\end{array}\right],
	\label{M-def}\\
	&\bS(k)\left[\begin{array}{c}
    	A_-(k)\\ B_+(k)\end{array}\right]=
   	 \left[\begin{array}{c}
    	A_+(k)\\ B_-(k)\end{array}\right].
	\label{S-def}
	\end{align}
the transfer and scattering matrices of the scattering system. If (\ref{we}) is nonlinear, their entries,
$M_{ij}(k)$ and $S_{ij}(k)$, are respectively nonlinear functions of $(A_-,B_-)$ and $(A_-,B_+)$. In the following we focus our attention to scattering phenomena defined by linear wave equations.\footnote{A linear wave equation is an equation of the form (\ref{we}) such that the linear combinations of its solutions are also solutions of this equation.}

Because  $(A_-,B_-)$ and $(A_+,B_+)$ determine the behavior of the solutions $\psi(x)$ at $x=-\infty$ and $x=+\infty$, the global existence and uniqueness of the solution of the initial-value problem defined by (\ref{we}) and (\ref{pw-m}) implies that $\bM(k)$ is an invertible matrix, i.e.,
	\be
	\det\bM(k)\neq 0.
	\label{det-M}
	\ee
Under this condition the scattering problem for the wave equation~(\ref{we}) is well-posed. We  therefore assume that it holds true. The inverse of $\bM(k)$ allows us to specify the asymptotic expression for the solutions of (\ref{we})  at $x=-\infty$ in terms of their asymptotic expression at $x=+\infty$. 

Let $\psi_\pm(k,x)$ be the solutions of (\ref{we}) that satisfy
	\be
	\psi_\pm(k,x)=e^{\pm ik x}~~~{\rm for}~~~x\to\pm\infty.
	\label{jost-1}
	\ee
Then Eq.~(\ref{M-def}) implies
	\bea
	\psi_-(k,x)&\to& M_{22}e^{ikx}+M_{12}(k)e^{-ikx}~~~{\rm for}~~~x\to+\infty,
	\label{jost-m}\\
	\psi_+(k,x)&\to&\frac{-M_{21}(k)e^{ikx}+M_{22}e^{-ikx}}{\det\bM(k)}
	~~~{\rm for}~~~x\to-\infty.
	\label{jost-p}
	\eea
$\psi_\pm$ are called the Jost solutions of the wave equation~(\ref{we}). Comparing (\ref{jost-1}) --
(\ref{jost-p}) with (\ref{psi-L}) and (\ref{psi-R}) and using the linearity of  (\ref{we}), we can respectively identify $\psi_l(k,x)$ and $\psi_r(k,x)$ with 
$\cN_+(k)T^l(k)\psi_+(k,x)$ and $\cN_-(k)T^r(k)\psi_-(k,x)$. Furthermore, this identification implies
	\begin{align}
	&M_{11}(k)=\frac{D(k)}{{T}^r(k)},
	&&M_{12}(k)=\frac{{R}^r(k)}{{T}^r(k)},
	&&M_{21}(k)=-\frac{{R}^l(k)}{{T}^r(k)},
	&& M_{22}(k)=\frac{1}{{T}^r(k)},
    	\label{Mij=}\\
	&R^l(k)=-\frac{M_{21}(k)}{M_{22}(k)}, &&T^l(k)=\frac{\det\bM(k)}{M_{22}(k)},
    	&&R^r(k)=\frac{M_{12}(k)}{M_{22}(k)}, &&T^r(k)=\frac{1}{M_{22}(k)},
         \label{RT=}
    	\end{align}
where
	\be
	D(k):=T^l(k)T^r(k)-R^l(k)R^r(k)=\frac{M_{11}(k)}{M_{22}(k)}.
	\label{D-def}
	\ee
	
We can similarly relate the entries of the scattering matrix to the reflection and transmission amplitudes by enforcing (\ref{S-def}) for the coefficient functions of the Jost solutions $\psi_\pm(k,x)$. In view of (\ref{jost-1}) -- (\ref{jost-p}), this gives 
	\begin{align}
	&S_{11}(k)=T^l(k), && S_{12}(k)=R^r(k), &&S_{21}(k)=R^l(k),&&S_{22}(k)=T^l(k).
	\label{Sij=}
	\end{align}
In particular, 
	\be
	\det\bS(k)=D(k).
	\label{det-S-D}
	\ee

The above-mentioned requirements on the global existence of the solutions of (\ref{we}) that satisfy asymptotic boundary conditions (\ref{pw-m}), (\ref{pw-p}), and (\ref{jost-1}) restrict the wave operator $\sW$. For example if $\sW$ is the Schr\"odinger operator $-\partial_x^2+v(x)$ for a potential $v:\R\to\C$, we can satisfy these requirements provided that $v(x)$ fulfills the Faddeev condition \cite{kemp}:
	\be
	\int_{-\infty}^\infty (1+|x|)|v(x)|dx<\infty.
	\label{F-condi}
	\ee
	

\section{Generalized unitarity relation}

Let us make the $k$-dependence of the solutions of the wave equation~(\ref{we}) explicit by using $\psi(k,x)$ in place of $\psi(x)$ in (\ref{pw-m}) and (\ref{pw-p}). Consider the implications of the transformations:
	\begin{align}
	&\psi(k,x)\stackrel{\cR}{\longrightarrow}\breve\psi(k,x):=(\cR\psi)(k,x):=\psi(-k,x), 
	\label{cR}\\
	&\psi(k,x)\stackrel{\cP}{\longrightarrow}\widetilde\psi(k,x):=(\cP\psi)(k,x):=\psi(k,-x),
	\label{cP}\\
	&\psi(k,x)\stackrel{\cT}{\longrightarrow}\overline\psi(k,x):=(\cT\psi)(k,x):=\psi(k,x)^*,
	\label{cT}\\
	&\psi(k,x)\stackrel{\cP\cT}{\longrightarrow}\widetilde{\overline\psi}(k,x):=
	(\cP\cT\psi)(k,x):=\psi(k,-x)^*.
	\label{cPT}
	\end{align}
It is not difficult to see that the transformed wave functions, $\breve\psi(k,x), \widetilde\psi(k,x)$, $\overline\psi(k,x)$, and $\widetilde{\overline\psi}(k,x)$ also tend to plane waves at spatial infinities. Therefore they determine scattering phenomena. By analogy to the definition of the transfer matrix $\bM(k)$ for $\psi(k,x)$, i.e., (\ref{M-def}), we can introduce the transfer matrices for $\breve\psi(k,x), \widetilde\psi(k,x)$, $\overline\psi(k,x)$, and $\widetilde{\overline\psi}(k,x)$. We respectively label them by $\bM(-k), \widetilde\bM(k)$, $\overline\bM(k)$, and $\widetilde{\overline\bM}(k)$. In view of (\ref{cR}) -- (\ref{cT}), we can show that
	\begin{align}
	&\bM(-k)=\bsigma_1\bM(k)\bsigma_1,
	&&\widetilde\bM(k)=\bsigma_1\bM(k)^{-1}\bsigma_1,
	\label{M-trans-1}\\
	&\overline\bM(k)=\bsigma_1\bM(k)^*\bsigma_1,
	&&\widetilde{\overline\bM}(k)=\bM(k)^{-1*},
	\label{M-trans-2}
	\end{align}
where $\bsigma_1$ is the first Pauli matrix;
	\[\bsigma_1:=\left[\begin{array}{cc} 0 & 1 \\ 1 & 0\end{array}\right].\]
	
Similarly we can introduce the reflection and transmission amplitudes for $\breve\psi(k,x), \widetilde\psi(k,x)$, $\overline\psi(k,x)$, and $\widetilde{\overline\psi}(k,x)$, which by virtue of their relationship to $\bM(-k), \widetilde\bM(k)$, $\overline\bM(k)$, and $\widetilde{\overline\bM}(k)$ and Eqs.~(\ref{M-trans-1}) and (\ref{M-trans-2}), take the form:
	\begin{align}
    	& R^l(-k)=-\frac{R^r(k)}{D(k)},
    	&&T^l(-k)=\frac{T^l(k)}{D(k)},
    	&&R^r(-k)=-\frac{R^l(k)}{D(k)},
    	&&T^r(-k)=\frac{T^r(k)}{D(k)},
    	\label{am-btrvr-rt}\\[6pt]
	& \widetilde R^l(k)= R^r(k) ,
    	&&\widetilde T^l(k)=T^r(k),
    	&&\widetilde R^r(k)=R^l(k),
    	&&\widetilde T^r(k)=T^l(k),
    	\label{am-P-trans-ampl}\\[6pt]
    	& \overline R^l(k)=-\frac{R^{r}(k)^*}{D(k)^*},
    	&&\overline T^l(k)=\frac{T^{l}(k)^*}{D(k)^*},
    	&&\overline R^r(k)=-\frac{R^{l}(k)^*}{D(k)^*},
    	&&\overline T^r(k)=\frac{T^{r}(k)^*}{D(k)^*},
	\label{am-T-trans-ampl}\\[6pt]
	&\widetilde{\overline R}^{l}(k)=-\frac{R^{l}(k)^*}{D(k)^*},
    	&&\widetilde{\overline T}^{l}(k)=\frac{T^{r}(k)^*}{D(k)^*},
    	&&\widetilde{\overline R}^{r}(k)=-\frac{R^{r}(k)^*}{D(k)^*},
    	&&\widetilde{\overline T}^{r}(k)=\frac{T^{l}(k)^*}{D(k)^*},
    	\label{am-PT-rt-transform}
    	\end{align}
respectively.

Next, we invert (\ref{am-btrvr-rt}) to express $R^{r}(k)$ and $T^{r}(k)$ in terms of $R^{l}(-k)$, $T^{r}(-k)$, and $D(k)$. Substituting the result in (\ref{D-def}), we find  
	\be
	D(k)\left[T^r(-k)T^l(k)+R^l(-k)R^l(k)-1\right]=0.
	\label{gen-rel-1}
	\ee
Similarly, we can solve (\ref{am-btrvr-rt}) for $R^{l}(k)$ and $T^{l}(k)$ in terms of $R^{r}(-k)$, $T^{l}(-k)$, and $D(k)$, and use (\ref{D-def}) to establish:
	\be
	D(k)\left[T^l(-k)T^r(k)+R^r(-k)R^r(k)-1\right]=0.
	\label{gen-rel-2}
	\ee
Equations (\ref{gen-rel-1}) and (\ref{gen-rel-1}) imply that whenever $D(k)\neq 0$,
	\be
	T^{l/r}(-k)T^{r/l}(k)+R^{l/r}(-k)R^{l/r}(k)=1.
	\label{gen-rel}
	\ee
This is a generalized unitarity relation that reduces to (\ref{id-1}) whenever the scattering system has reciprocal transmission and $D(k)\neq 0$ for all $k\in\R^+$. Both of these conditions are satisfied for scattering systems determined by the Schr\"odinger equation for a local time-reversal invariant (real) or $\cP\cT$-symmetric potential. According to the reciprocity theorem they have reciprocal transmission, and as we show in the sequel they satisfy $|D(k)|=1$. To see this, first we note that according to (\ref{M-trans-2}) the transfer matrix for time-reversal-invariant and $\cP\cT$-symmetric systems\footnote{By definition, time-reversal-invariance and $\cP\cT$-symmetry of a scattering system respectively mean that its reflection and transmission amplitudes, and consequently its transfer and scattering matrices are invariant under time-reversal and $\cP\cT$ transformations.} respectively fulfil
	\bea
	\bM(k)^*&=&\bsigma_1\bM\,\bsigma_1,
	\label{M-T-sym}\\
	\bM(k)^*&=&\bM(k)^{-1}.
	\label{M-PT-sym}
	\eea
We can use these equations to show that
	\begin{align}
	\mbox{$\cT$-symmetry}&\quad\xto\quad
	M_{11}(k)^*=M_{22}(k),
	\label{T-sum-01}\\
	\mbox{$\cP\cT$-symmetry}&\quad\xto\quad
	M_{11}(k)^*=\frac{M_{22}(k)}{\det\bM(k)}.
	\label{PT-sum-01}
	\end{align}
For time-reversal-invariant systems, Eqs.~(\ref{D-def}) and (\ref{T-sum-01}) imply: 
	\be
	|D(k)|=\left|\frac{M_{11}(k)}{M_{22}(k)}\right|=
	\left|\frac{M_{11}(k)^*}{M_{22}(k)}\right|=1.
	\ee
In light of (\ref{D-def}) and (\ref{PT-sum-01}), we also find the following result for $\cP\cT$-symmetric scattering systems.
	\be
	|D(k)|=\left|\frac{M_{11}(k)}{M_{22}(k)}\right|=
	\left|\frac{M_{11}(k)}{\det\bM(k)}\right|\left|\frac{\det\bM(k)}{M_{22}(k)}\right|=
	\left|\frac{M_{22}(k)^*}{M_{11}(k)^*}\right|=
	\left|\frac{M_{22}(k)}{M_{11}(k)}\right|=\frac{1}{|D(k)|},
	\ee
which means $|D(k)|=1$. 

Note that the proof of the identity $|D(k)|=1$ we have just presented does not make use of the transmission reciprocity. Therefore it holds for every scattering system possessing time-reversal invariance or $\cP\cT$-symmetry. In view of (\ref{gen-rel-2}), it implies that the reflection and transmission amplitudes of these systems fulfill (\ref{gen-rel}) for all $k\in\R^+$.

For scattering systems that are neither time-reversal-invariant nor $\cP\cT$-symmetric, there may exist values of $k$ for which $D(k)=0$, in which case (\ref{gen-rel}) may be violated for these values of $k$.
According to (\ref{det-S-D}), these are the real and positive zeros $k_0$ of $\det\bS(k)$. Clearly  $\det\bS(k_0)=0$ means that $\bS(k_0)$ has a vanishing eigenvalue, i.e., there are complex numbers $A_{0-}$ and $B_{0+}$ such that
	\be 
	\bS(k_0)\left[\begin{array}{c} A_{0-}\\ B_{0+}\end{array}\right]=
	\left[\begin{array}{c} 0\\ 0\end{array}\right].
	\ee
In light of (\ref{pw-m}), (\ref{pw-p}), and (\ref{S-def}), this equation proves the existence of a solution $\psi_{\rm in}(k,x)$ of the wave equation that satisfies purely incoming asymptotic boundary conditions for $k=k_0$, i.e.,
	\[\psi_{\rm in}(k_0,x)\to \left\{
	\begin{array}{ccc}
	A_{0-}e^{ik_0 x} & {\rm for} & x\to-\infty,\\
	B_{0+}e^{-ik_0 x} & {\rm for} & x\to+\infty.\end{array}\right.\]
This solution describes a rather remarkable situation where the system absorbs a pair of incident waves traveling towards it in opposite directions. This phenomenon is called coherent perfect absorption or antilasing \cite{CPA1,longhi,longhi-PT-CPA,CPA2,jpa-2012}.

The above analysis shows that for every scattering system and $k\in\R^+$, either $k$ is a wavenumber at 
which the system acts as a coherent perfect absorber or its reflection and transmission amplitudes satisfy the generalized unitarity relation (\ref{gen-rel}). 

Let us conclude by noting that the term `generalized unitarity relation' refers to the fact that for a real scattering potential where the wave operator is a Hermitian Schr\"odinger operator, this relation reduces to the unitarity relation (\ref{unitarity}). This follows from the reciprocity theorem and Eqs.~(\ref{am-btrvr-rt}) and (\ref{am-T-trans-ampl}), which for time-reversal-invariant systems  imply
	\begin{align*}
	&R^{l/r}(-k)=R^{l/r}(k)^*, && T^{l/r}(-k)=T^{l/r}(k)^*.
	\end{align*}
	
\subsection*{Acknowledgments}

This work has been supported by the Scientific and Technological Research Council of Turkey (T\"UB\.{I}TAK) in the framework of the project no: 114F357 and by Turkish Academy of Sciences (T\"UBA).

\ed
\begin{thebibliography}{99}

\bibitem{jpa-2014c} A.~Mostafazadeh, Generalized unitarity and reciprocity relations for $\cP\cT$-symmetric scattering potentials, J.~Phys.~A: Math.\ Theor.\ {\bf 47},  505303 (2014).

\bibitem{ahmed} Z.~Ahmed, Schr\"odinger transmission through one-dimensional complex potentials, Phys.\ Rev.~A {\bf 64}, 042716 (2001)

\bibitem{jpa-2009} A.~Mostafazadeh and H.~Mehri-Dehnavi, Spectral singularities, biorthonormal systems and a two-parameter family of complex point interactions, J.~Phys.~A: Math.\ Theor.\ {\bf 42}, 125303 (2009).

\bibitem{landau-lifshitz} L.~D.~Landau and E.~M.~Lifshitz, {\rm Quantum Mechanics}, Pergamon, Oxford, 1977.


\bibitem{lin} Z.~Lin, H.~Ramezani, T.~Eichelkraut, T.~Kottos, H.~Cao, D.~N.~Christodoulides, Unidirectional invisibility induced by PT-symmetric periodic structures, Phys.\ Rev.\ Lett.~{\bf 106}, 213901 (2011).

\bibitem{pra-2013a} A.~Mostafazadeh, Invisibility and $\cP\cT$-symmetry, Phys.\ Rev.~A {\bf 87}, 012103 (2013).

\bibitem{ss-review} A.~Mostafazadeh, Physics of Spectral Singularities, in Proceedings of XXXIII Workshop on Geometric Methods in Physics, held in Bialowieza, Poland, June 29-July 5, 2014, Trends in Mathematics, pp.~145-165, Springer International Publishing, Switzerland, 2015; preprint arXiv:1412.0454.

\bibitem{konotop-review} V.~V.~Konotop, J.~Yang, and D.~A.~Zezyulin, Nonlinear waves in PT -symmetric systems, Rev.\ Mod.\ Phys.\ {\bf 88}, 035002 (2016).

\bibitem{ge} L.~Ge, Y.~D.~Chong, and A,~D.~Stone, Conservation relations and anisotropic transmission resonances in one-dimensional $\cP\cT$-symmetric photonic heterostructures, Phys.\ Rev.~A {\bf 85} 023802 (2012).

\bibitem{ahmed-2012} Z.~Ahmed, New features of scattering from a one-dimensional non-Hermitian (complex) potential, J.~Phys.~A: Math.\ Theor.\ {\bf 45},  032004 (2012).

\bibitem{muga} J.~G.~Muga, J.~P.~Palao, B.~Navarro, and I.~L.~Egusquiza, Complex absorbing potentials, Phys.\ Rep.~{\bf 395}, 357-426 (2004).

\bibitem{muga2}  A.~Ruschhaupt, T.~Dowdall, M.~A.~Simon, and J.~G.~Muga, Asymmetric scattering by non-hermitian potentials, preprint arXiv 1709:07027.

\bibitem{kemp} R.~R.~D.~Kemp, A singular boundary value problem for a non-self-adjoint differential opeartor, Canadian J.~Math. \textbf{10}, 447-462 (1958).

\bibitem{CPA1}  Y.~D.~Chong, L.~Ge, H.~Cao, and A.~D.~Stone, Coherent perfect absorbers: Time-reversed lasers, Phys.\ Rev.\ Lett.\ {\bf 105}, 053901 (2010).

\bibitem{longhi} S.~Longhi, Backward lasing yields a perfect absorber, Physics {\bf 3}, 61 (2010).

\bibitem{longhi-PT-CPA} S.~Longhi, $\cP\cT$-symmetric laser absorber, Phys. Rev. A {\bf 82}, 031801 (2010).

\bibitem{CPA2} W.~Wan, Y.~Chong, L.~Ge, H.~Noh, A.~D.~Stone, and H.~Cao,  Time-reversed lasing and interferometric control of absorption, Science {\bf 331}, 889-892 (2011).

\bibitem{jpa-2012} A.~Mostafazadeh, Self-dual spectral singularities and coherent perfect
absorbing lasers without $\cP\cT$-symmetry, J.~Phys.~A: Math.\ Gen.~{\bf 45}, 444024 (2012).





\end{thebibliography}
